\documentclass[final]{raa06}           
\usepackage{graphicx,times}             
\usepackage{natbib}
\usepackage{amssymb,amsmath}
\bibpunct{(}{)}{;}{a}{}{,}
\usepackage[colorlinks=true, citecolor=blue]{hyperref}%

\usepackage{graphicx,kantlipsum,setspace}
\usepackage{caption}
\usepackage{graphics,epsf}
\usepackage{amsmath}                
\usepackage{amsfonts}               
\usepackage{amssymb}                
\usepackage{epsfig}                 
\usepackage{appendix}
\usepackage{graphicx}
\usepackage{float}
\usepackage{color}
\usepackage{multirow}
\usepackage{colortbl}
\usepackage[para,online,flushleft]{threeparttable}
\usepackage{xcolor}

\hypersetup{citecolor=blue, 
            linkcolor=red, 
            menucolor=blue, 
            urlcolor=blue}  

\newcommand{\cm}{{~\rm cm}}

\newcommand{\km}{{~\rm km}}
\newcommand{\s}{{~\rm s}}

\newcommand{\g}{{~\rm g}}

\newcommand{\erg}{{~\rm erg}}

\newcommand{\kpc}{{~\rm kpc}}

\begin{document}

   \title{Predicting gravitational waves from jittering-jets-driven core collapse supernovae
}

   \volnopage{Vol.0 (20xx) No.0, 000--000}      
   \setcounter{page}{1}          

   \author{Noam Soker
    }

   \institute{Department of Physics, Technion, Haifa, 3200003, Israel;   {\it   soker@physics.technion.ac.il}\\
\vs\no
   {\small Received~~20xx month day; accepted~~20xx~~month day}}

\abstract{I estimate the frequencies of gravitational waves from jittering jets that explode core collapse supernovae (CCSNe) to crudely be 5-30 Hz, and with  strains that might allow detection of Galactic CCSNe. The jittering jets explosion mechanism (JJEM) asserts that most CCSNe are exploded by jittering jets that the newly born neutron star (NS) launches within few seconds. According to the JJEM, instabilities in the accreted gas lead to the formation of intermittent accretion disks that launch the jittering jets. Earlier studies that did not include jets calculated the gravitational frequencies that instabilities around the NS emit to have a peak in the crude frequency range of 100-2000 Hz. Based on a recent study, I take the source of the gravitational waves of jittering jets to be the turbulent bubbles (cocoons) that the jets inflate as they interact with the outer layers of the core of the star at thousands of kilometres from the NS. The lower frequencies and larger strain than those of gravitational waves from instabilities in CCSNe allow future, and maybe present, detectors to identify the gravitational wave signals of jittering jets. Detection of gravitational waves from local CCSNe might distinguish between the neutrino-driven explosion mechanism and the JJEM.   
}

 \authorrunning{N. Soker}            
\titlerunning{Gravitational waves from jittering jets CCSNe}  
   
      \maketitle

\section{Introduction} 
\label{sec:intro}

Since the early days of gravitational wave detectors core collapse supernovae (CCSNe) have been considered as potential sources of gravitational waves (e.g., \citealt{deFreitasPacheco2010}), with intensified research in recent years (e.g., \citealt{AfleBrown2021, Gilletal2022, SaizPerezetal2022}).
Gravitational waves from CCSNe are yet to be detected (e.g., \citealt{Szczepanczyketal2023}). Recent studies concentrate on the expected gravitational waves from CCSNe during the explosion process (e.g., \citealt{PowellMuller2019, Linetal2023, Mezzacappaetal2023, Wolfeetal2023, PastorMarcosetal2023}; for more on the results of some studies see section \ref{sec:Identification}) and shortly after explosion, e.g., in relation to magnetar formation (e.g., \citealt{Chengetal2023, Menonetal2023}).  

Studies that calculate the properties of  gravitational waves from CCSN explosions ignore the role of jittering jets. The goal of this exploratory study is to estimate the expected contribution of jittering jets to gravitational wave emission from CCSNe.  
The motivation for this study are results from recent years that support the jittering jets explosion mechanism (JJEM) of CCSNe (e.g., \citealt{Soker2022Rev, Soker2022SNR0540, Soker2023Classes, ShishkinSoker2023}), and the very recent study by \cite{Gottliebetal2023} who found that the turbulent cocoons that energetic relativistic jets form (e.g., \citealt{Izzoetal2019}) can be a strong source of gravitational waves. A cocoon is the convective bubble that a jet inflates and is filled with the shocked jet's material and the shocked ambient material. 
\cite{Gottliebetal2023} simulated relativistic and very energetic jets, $\approx 10^{52}-10^{53} \erg$, that are relevant to rare CCSNe where the pre-explosion core is rapidly rotating and the collapsing core is likely to form a black hole. There are many studies of such rare CCSNe (but not of the gravitational waves from jets) that have fixed-axis jets (e.g., \citealt{Khokhlovetal1999, Aloyetal2000,  Maedaetal2012, LopezCamaraetal2013, BrombergTchekhovskoy2016,  Nishimuraetal2017, WangWangDai2019RAA, Grimmettetal2021, Perleyetal2021, Gottliebetal2022, ObergaulingerReichert2023}).

In this study, however, I deal with non-relativistic jets where each jet-pair has a much lower-energy of $\approx 10^{50} \erg$. The JJEM asserts that such jets explode most CCSNe (e.g., \citealt{Soker2010, PapishSoker2011,  Soker2020RAA, ShishkinSoker2021, Soker2023gap}).
The newly-born neutron star (NS), or in some cases a black hole, launches the jets as it accretes mass through an accretion disk. There are two sources of the angular momentum of the accretion disk (e.g., \citealt{Soker2023gap}). These are pre-collapse core rotation that has a fixed angular momentum axis, and the convective motion in the pre-collapse core (e.g., \citealt{PapishSoker2014Planar, GilkisSoker2015, Soker2019SASI, ShishkinSoker2022}) or envelope (e.g., \citealt{ Quataertetal2019, AntoniQuataert2022, AntoniQuataert2023}) that has a stochastically varying angular momentum axis. When the pre-collapse core angular momentum is low the accretion disk has rapidly varying axis direction. Each accretion episode through a given accretion disk lasts for a limited period of time and leads to one jet-launching episode of two opposite jets.    
 
The convective fluctuations serve as seed perturbations that are amplified by instabilities behind the stalled shock, which is at $\simeq 100-150 \km$ from the newly born NS. Namely, the same instabilities that give rise to gravitational waves in the frame of the neutrino-driven explosion mechanism (e.g., \citealt{Mezzacappaetal2020}), which does not include jets, exist also in the JJEM. The JJEM has in addition the jittering jets that inflate turbulent bubbles (cocoons) that might emit gravitational waves according to the new results of \cite{Gottliebetal2023}. 

In the present, still exploratory, study I present the first prediction, although very crude, for gravitational waves in the frame of the JJEM. I do this by appropriately scaling the recent results that \cite{Gottliebetal2023} obtained for gravitational waves from much more energetic jets than the jittering jets (section \ref{sec:GravitaionalWaves}). I then present the general characteristic of the strain of JJEM-driven CCSNe (section \ref{sec:Identification}). I summarize the results (section \ref{sec:Summary}) and strongly encourage simulations of gravitational waves from jittering jets in CCSNe.

\section{Estimating gravitational waves from jittering jets} 
\label{sec:GravitaionalWaves}

The calculation of gravitational waves by CCSNe as expected in the JJEM requires very demanding three-dimensional hydrodynamical simulations. In this preliminary study I make crude estimates by scaling the results of \cite{Gottliebetal2023} who conduct simulations of long-lived relativistic jets with energies of $\simeq 10^{52} - 10^{53} \erg$. 

In the JJEM the jets are relatively short-lived and have a typical velocity of $0.3-0.5 c$ (e.g., \citealt{PapishSoker2014a}; indeed, \citealt{Guettaetal2020} claim that neutrino observations limit the jets in most cases to be non-relativistic). In an explosion process there are $\approx {\rm few} - 30$ jet-launching episodes, with a typical activity time of each episode of $\simeq 0.01-0.1 \s$, and a typical energy of the two jets of  $\approx 10^{50} \erg$ to ${\rm few} \times 10^{50} \erg$ \citep{PapishSoker2014a}.  

\cite{Gottliebetal2023} estimate the range of frequencies of the gravitational waves when the jets' axis is at a large angle to the line of sight (off-axis) to be between $f_{\rm min} \simeq 1/ \Delta t_{\rm jc}$ and $f_{\rm max} \simeq c_s/\Delta r_{\rm sh}$, where $\Delta t_{\rm jc}$ is the time the jets energize  the cocoons, $c_s$ is the sound speed, and $\Delta r_{\rm sh}$ is the width of the shell formed by the shock. For their simulations this range is $\approx 0.1-2000 {~\rm Hz}$. The on-axis emission, i.e., when the jets' axis is at a very small angle to the line of sight, has a strain amplitude that is more than an order of magnitude smaller than for the off-axis emission and the strain amplitude peaks at frequencies of $10-100 {~\rm Hz}$. 

\begin{figure} 
\centering
\includegraphics[trim=0.2cm 2.75cm 5.0cm 4.0cm ,clip, scale=0.67]{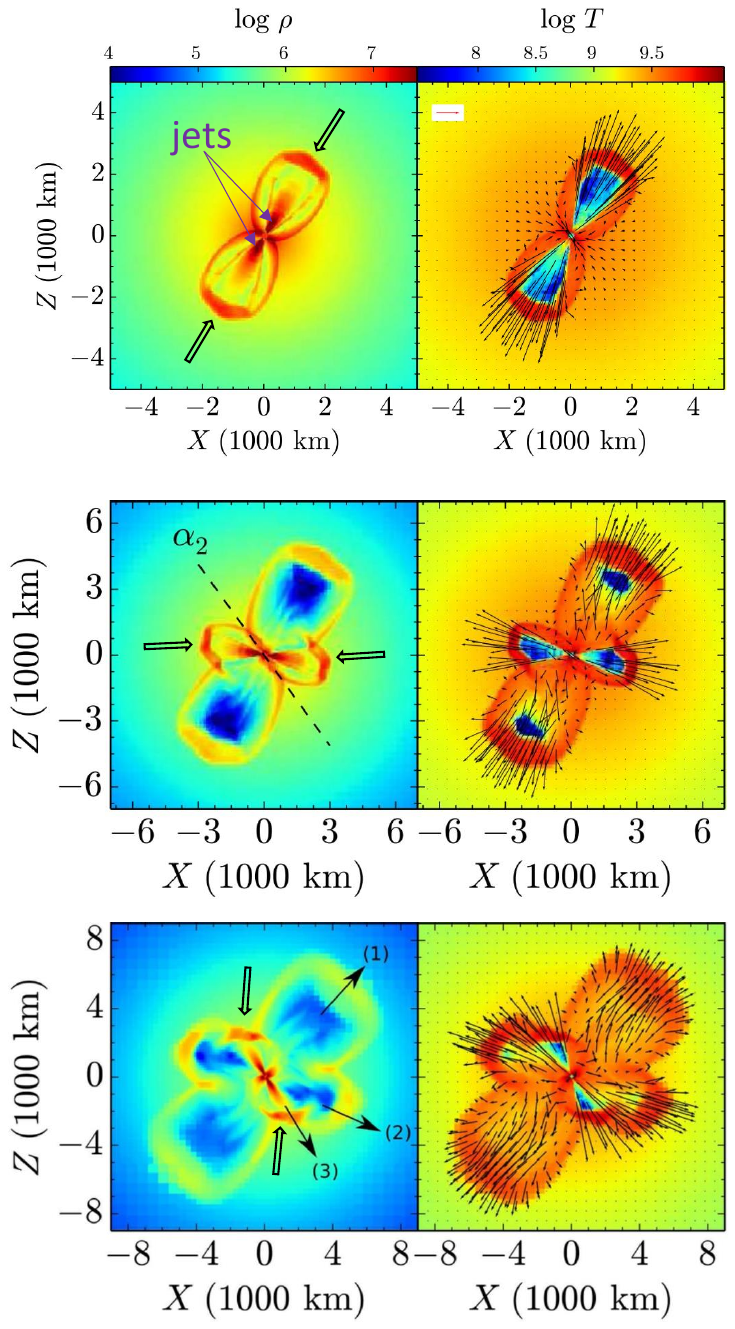}  
\caption{Density (left column with a colour coding in logarithmic scale and units of $\g \cm^{-3}$) and temperature (right column in log scale in units of K) maps at three times of three-dimensional hydrodynamical simulation of jittering jets taken from \cite{PapishSoker2014Planar}.
There are three jet-launching episodes, each composed of two opposite jets, one episode after the other with activity times of $0-0.05 \s$ direction 1 in the lower panel, $0.05-0.1 \s$ direction 2, and $0.1-0.15 \s$ direction 3. 
I added double-lined arrows to point at the two opposite masses at the cocoon (bubble) head. While the first jet-pair inflates axisymmetric cocoons, the following cocoons largely deviate from axisymmetry. Velocity is proportional to the arrow length on the right column, with inset showing an arrow for $30,000 \km \s^{-1}$.
}
\label{Fig:Papish1}
\end{figure}

To scale for one pair of jittering jets I consider the three-dimensional simulations by \cite{PapishSoker2014Planar}. They simulated three pairs of jittering jets that have their axes on the same plane, each jet-launching episode lasting for $0.05 \s$. In Figure \ref{Fig:Papish1} I present the density and temperature maps in the jittering plane of these jets. 
In each jet-launching episode the two opposite jets are seen as two opposite high density (red color on the left column) strips touching the center. While the first jet-pair inflates axisymmetric cocoons (bubbles), the second and third jet-pairs inflate non-axisymmetric bubbles. This is seen by the compressed gas at the head of the cocoon (bubble) that I point at with the double-lined arrows.

From the density maps of Figure \ref{Fig:Papish1} I estimate the width of the shells of the bubbles/cocoons that the jets inflate to be $\Delta r_{\rm sh} \simeq 200-500 \km$, and from the temperature maps the sound speed is $c_s \simeq 5000 \km \s^{-1}$.  This gives with the definition of \cite{Gottliebetal2023} $f_{\rm max} \simeq c_s/\Delta r_{\rm sh} \approx 10-25 {~\rm Hz}$. In the JJEM the typical duration of a jet-launching episode is $\Delta t_{\rm j} \approx 0.01-0.1 \s$. Even if the jets lasts for $\simeq 0.01 \s$, the interaction with the core material lasts longer. For that the interaction time is more likely to be $\Delta t_{\rm jc} \approx 0.05-0.2 \s$, which gives with the definition of \cite{Gottliebetal2023} $f_{\rm min} \simeq 1/ \Delta t_{\rm jc} \approx 5-20 {~\rm Hz}$ for typical jittering jets, but with large uncertainties. The short-duration jets will have small energy and therefore small strain amplitude. The longer-duration jets have more energy. Therefore, waves with lower frequency are more likely to be detected, i.e., $f_{\rm min} \simeq 5-10 {~\rm Hz}$. 

The relatively small ratio of $f_{\rm max}/f_{\rm min} \approx 1-5$ that I find here shows that the typical spectrum of the gravitational waves of jittering jets is qualitatively different from the case that \cite{Gottliebetal2023} study. 
In the case of the JJEM I expect the spectrum to be in narrow range of 
\begin{equation}
f_{\rm JJEM} \approx 5-30 {~\rm Hz}.  
\label{eq:fJJEM}
\end{equation}

As seen in Fig. \ref{Fig:Papish1} the size of the cocoon is smaller than the typical wavelength of $\approx 20,000 \km$, which makes phase cancellation very small. 

Scaling equation (2) of \cite{Gottliebetal2023} for the strain amplitude for one pair of jets out of many pairs in the JJEM, gives   
\begin{equation}
h \approx 4 \times 10^{-22} 
\left( \frac{D}{10 \kpc} \right)^{-1}
\left( \frac{E_{\rm 2j}}{10^{50 \erg}} \right). 
\label{eq:Strain}
\end{equation}
I also consider the following quantity that is used in the study of gravitational waves from CCSNe
\begin{equation}
\begin{split}
\frac{h}{\sqrt{f}} \approx  10^{-22} &
\left( \frac{D}{10 \kpc} \right)^{-1}
\left( \frac{E_{\rm 2j}}{10^{50 \erg}} \right)  \\
\times & 
\left( \frac{f_{\rm JJEM}}{15~{\rm Hz}} \right)^{-1/2} {\rm Hz}^{-1/2} ,
\label{eq:StrainDf}
\end{split}
\end{equation}
were I scaled with the expected frequency range for jittering jets from equation (\ref{eq:fJJEM}).

I note that equations (\ref{eq:Strain}) and (\ref{eq:StrainDf}) treat each jet-launching episode as an independent event. If several episodes are considered to inflate only two opposite large bubbles (lower panel of Figure \ref{Fig:Papish1}) then the energy in the scaling of the equations should be the sum of several jet-launching episode. Namely, the scaling energy should be $\simeq {\rm few} \times 10^{50}$ leading to a strain larger by a factor of a few.  

\section{Identification of gravitational waves from jittering jets} 
\label{sec:Identification}

Several papers calculated the gravitational waves properties from CCSNe when jets are not included (e.g., \citealt{Radiceetal2019, Andresenetal2021, Mezzacappaetal2023}), i.e., in the frame of the delayed neutrino explosion mechanism (e.g.,   \citealt{BetheWilson1985, Hegeretal2003, Janka2012, Nordhausetal2012, Mulleretal2019Jittering, Fujibayashietal2021, Bocciolietal2022, Nakamuraetal2022, Olejaketal2022}).
\cite{Mezzacappaetal2020}, for example, find that low-frequency emission, $\lesssim 200 {~\rm Hz}$, is emitted by the neutrino-driven convection and the standing accretion shock instability in the gain layer behind the stalled shock, while high-frequency emission, $\lesssim 200 {~\rm Hz}$, is emitted by convection in the  proto–NS. These studies find that the emission is mainly at frequencies of $\approx 10 - 2000 {~\rm Hz}$ with larger strain amplitudes at frequencies of $\approx 100 - 1000 {~\rm Hz}$ (e.g., \citealt{Srivastavaetal2019}). 

\begin{figure} 
\centering
\includegraphics[trim=2.5cm 14.75cm 2.0cm 1.9cm ,clip, scale=0.51]{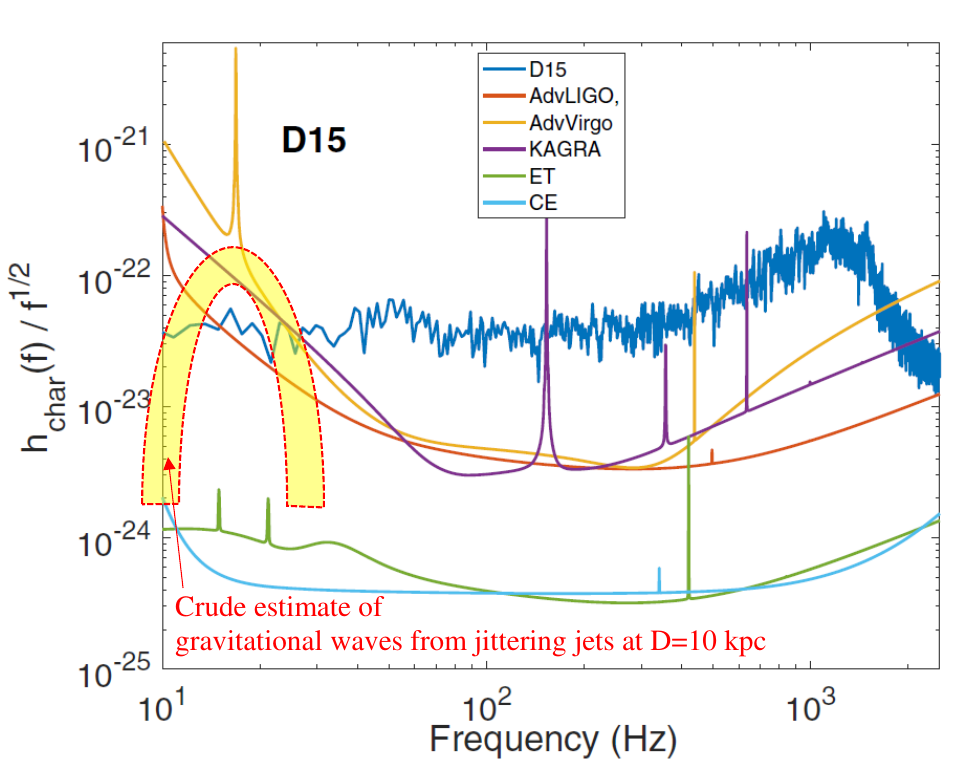}  
\caption{A figure from \cite{Mezzacappaetal2023} to which I added a crude estimate of the characteristic spectrum of $h f^{-1/2}$ from jittering jets in a CCSN at a distance of $D=10 \kpc$ (the horseshoe-shaped yellow zone). The signal in yellow is for one jet-launching episode. If several jet-launching episodes are considered to inflate only two opposite large bubbles (lower panel of Figure \ref{Fig:Papish1}) then the strain will be larger, as it is about the sum of these episodes. 
Other marks are as in the original figure.  
The blue line is the calculation by \cite{Mezzacappaetal2023} of the characteristic gravitational wave strain from a CCSN of a $15 M_\odot$ stellar model. The five other lines represents the sensitivity curves for gravitational wave
detectors: Advanced Laser Interferometer Gravitational Observatory
(AdvLIGO), Advanced VIRGO, and Kamioka Gravitational
Wave Detector (KAGRA) that are current-generation gravitational wave detectors, and the more sensitive next-generation detectors, Cosmic Explorer and Einstein Telescope. The predicted full gravitational wave spectrum includes both the contributions from the regions near the NS that exist both in the JJEM and in the neutrino-driven explosion mechanism (blue line), and the contribution of the jittering jets.   
} 
\label{Fig:StarinFrequency}
\end{figure}

The gain region and the convection in the proto-NS exist also in the JJEM. Neutrino heating play roles also in the JJEM \citep{Soker2022Boosting}. Therefore, the contributions of the gain region and the proto-NS to gravitational waves in the JJEM are similar to those in the delayed neutrino explosion mechanism. In the JJEM there is the additional contribution of the cocoons that the jets inflate in the core and envelope of the exploding star. In section \ref{sec:GravitaionalWaves} I crudely estimated this contribution for jittering jets interacting with the core of the exploding star. In Figure  \ref{Fig:StarinFrequency} I present results from \cite{Mezzacappaetal2023}. The results is of the characteristic gravitational wave strain from a CCSN in the frame of the delayed neutrino explosion mechanism of a $15 M_\odot$ stellar model. I added my crude estimate of a typical contribution of jittering jets (the horseshoe-shaped yellow region on the graph). 

\begin{figure*} 
\centering
\includegraphics[trim=0.5cm 3.2cm 2.0cm 6.9cm ,clip, scale=0.75]{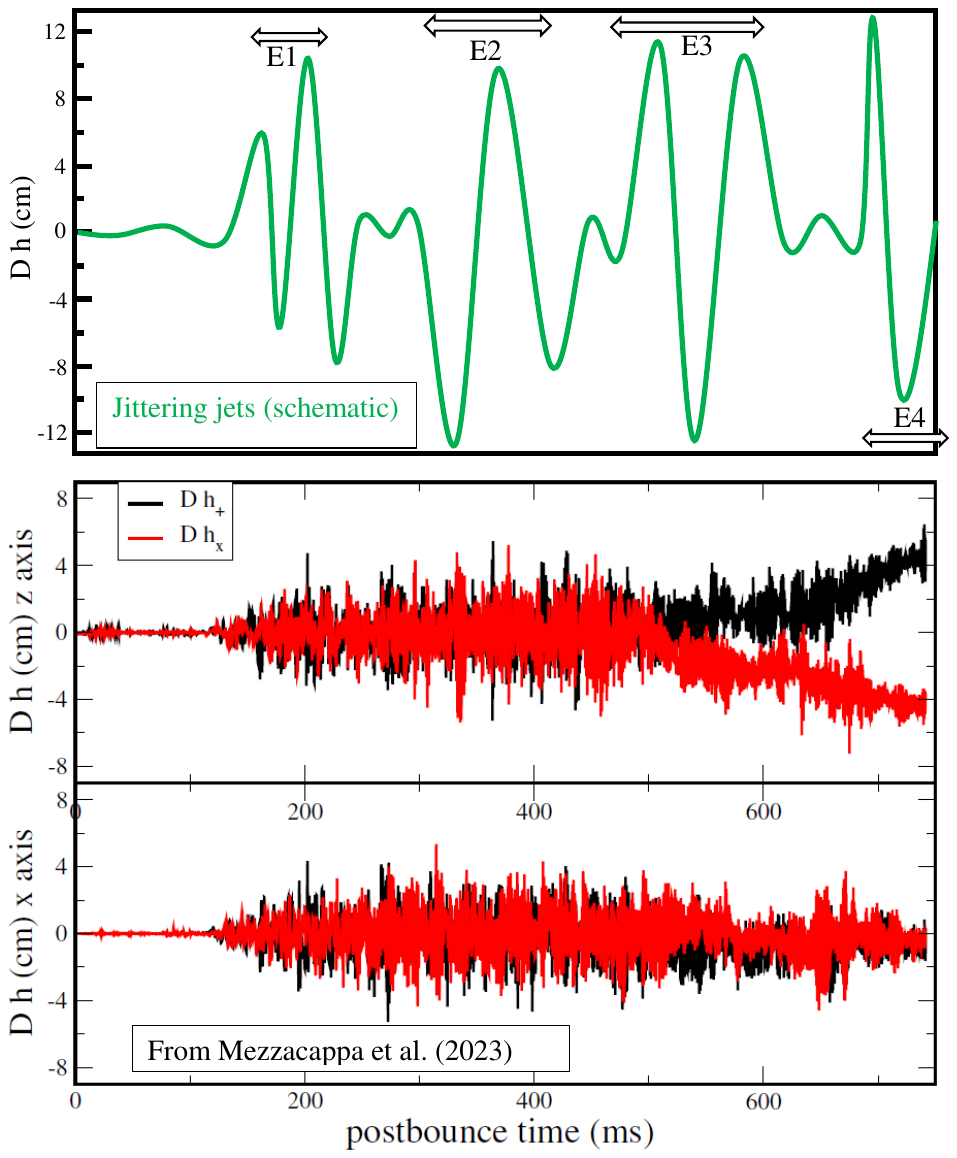}  
\caption{The gravitational wave strain times distance as function of time during the early explosion process. The upper panel is a schematic presentation of a possible wave form from jittering jets. The typical amplitude and frequency are according to equations (\ref{eq:Strain}) and (\ref{eq:fJJEM}), respectively. The double-headed arrows present the contributions of four jet-launching episodes, E1 to E4. 
The lower panel is from \cite{Mezzacappaetal2023} for calculations based on a simulation that do not include jets of an exploding stellar model of $15 M_\odot$. 
} 
\label{Fig:StarinTime}
\end{figure*}

The peak of the contribution of the jittering jets is at much lower frequencies than the peak of the other components of CCSNe. In addition, there will be variations with time as the jittering jets are active intermittently. As said, simulations of the JJEM are highly demanding because for the calculation of gravitational waves high-resolution simulations are need to resolve the convection in the cocoon and the head of the jet-core interaction. At this point I only present the possible schematic behavior of the strain as function of time due to the contribution of jittering jets. 
In the upper panel of Figure \ref{Fig:StarinTime} I schematically present such a gravitational wave signal due only to jittering jets. I describe the distance times the strain of four jet-launching episodes (but more are expected at later time until the star explodes). Over the time period $0.2 \s -0.7 \s$ the average frequency is $16 {~\rm Hz}$. 
As commonly done I take $t=0$ at the bounce of the shock wave from the newly born NS. There is some time delay until instabilities start to feed the intermittent accretion disks that launch the jets. These instabilities give rise to high-frequency-gravitational waves (e.g., \citealt{Radiceetal2019, Andresenetal2021}). In the lower panel of Figure \ref{Fig:StarinTime}  I present one figure from  \cite{Mezzacappaetal2023} that shows their calculation for the gravitational wave of a CCSN of $15 M_\odot$ stellar model. The expected signal is the sum of all contributions.   

My crude estimate of gravitational waves from jittering jets show that their signal is qualitatively different than that of the other components that are close to the NS, $\lesssim 100 \km$. The jittering jets add long period modulations to the short-period waves from the other components. For a nearby CCSN even the present Advanced LIGO detector might separate the signal of the jittering jets from the other components. This depends on the signal-to-noise ratio that should be calculated with future simulations of jittering jets. Future detectors will be able to do so for CCSNe in the local group. 

\section{Summary} 
\label{sec:Summary}

Based on the very recent results by \cite{Gottliebetal2023}, which I scaled from long-lasting energetic relativistic jets in super-energetic CCSN to short-lived low-energy non-relativistic jets in common CCSNe, I concluded that jittering jets lead to detectable gravitational wave signals. The source of the gravitational waves is the turbulence in the cocoons that the jets inflate (Figure \ref{Fig:Papish1}). 
Whether present detectors can reveal the gravitational wave signals of jittering jets depend on the signal-to-noise ratio that simulations of jittering jets should calculate, and of course on the distance to the CCSN. Future detectors will be able to reveal the jittering jets signal from CCSNe in the local group (Fig. \ref{Fig:StarinFrequency}). 

The frequencies of the expected gravitational wave signals from jittering jets are lower than the other components of CCSNe, as I mark by the yellow horseshoe-shaped region in Figure \ref{Fig:StarinFrequency}. I schematically present a gravitational wave signal from jittering jets in the upper panel of Figure \ref{Fig:StarinTime}, and compared it with calculations from a CCSN simulation that include no jets from \cite{Mezzacappaetal2023}. The signal from jittering jets can be clearly distinguished from the other gravitational waves sources in CCSNe (depending on signal-to-noise ratio and the distance of the CCSN).   

This, still exploratory, study calls for the performance of highly-demanding simulations of jittering jets and the calculation of their gravitational wave signals. The simulation must be of very high resolution as to resolve the turbulence in the cocoon. 

Because I expect jittering jets to explode most CCSNe, my prediction for the gravitational wave signals from nearby CCSNe differ from the prediction of studies that include no jets. 





\label{lastpage}

\end{document}